# Phonon origin of high $T_c$ in Superconducting cuprates

E.A.Mazur [1(a)], Yu.Kagan [2],

[1] *National research nuclear university "MEPhI", Kashirskoe shosse, 31, Moscow, 115409, Russia*

[2] *NRC Kurchatov institute, Akademika Kurchatova pl.1, Moscow, 123182, Russia*

**Abstract** Eliashberg theory (ET) generalized for the account of the peculiar properties of the finite zone width electron-phonon (EP) systems with the non constant electron density of states, the electron-hole nonequivalence, chemical potential renormalization with doping and frequency, and electron correlations in the vertex function is used for the study of $T_c$ in cuprates. The phonon contribution to the nodal anomalous electron Green function (GF) in cuprates is considered. The pairing on the full width of the electron zone was taken into account, not just on the Fermi surface. It is found that the finite zone width phenomenon in the newly derived Eliashberg equations for the finite zone width EP system together with the abrupt fall of the density of states above the Fermi surface are the crucial factors for the appearance of the high temperature superconductivity phenomenon. It is shown that near the optimal doping in the hole-doped cuprates high $T_c$ value is reproduced with the EP interaction constant obtained from tunnel experiments.

## 1 Eliashberg theory for the Finite Electron Zone Width with the Non Constant Density of the Electron States

It is believed that high $T_c$ value in the case of the EP mechanism of superconductivity is reproduced by Eliashberg theory (ET) (see reviews [1–6] and references therein) only with unreasonably high constants of the EP interaction $\lambda \geq 10$. In fact, at such high $\lambda \geq 3$ the different version of the theory of the EP systems [7] instead of the Migdal-Eliashberg theory should be applied. At the same time it was found that the real EP interaction constant $\lambda$ in the cuprates is not significantly exceeding unity $\lambda \geq 1$ [4]. The inability to explain the high $T_c$ value in the cuprates by using the standard ET has led to attempts to interpret the high $T_c$ value in the materials of this class as a display of new mechanisms of the electron pairing [1–6]. In [8,9] it was shown that the real part $\operatorname{Re}\Sigma$ and imaginary part $\operatorname{Im}\Sigma$ of the SE part renormalization are not restricted to the frequency value $\omega$ of the order of the limiting phonon frequency $\omega_D$ and spread to the significantly higher frequency value $\omega \gg \omega_D$. In the present paper it is shown that $T_c$ calculations within the advanced ET that takes into account the finiteness of the electron band and the abrupt change in the electron density of states near the Fermi surface well reproduce the high $T_c$ value in the cuprates. Such an effect was not represented in the earlier versions of the ET [1–6,10]. The vertex matrix (in $\hat{\tau}_i$ space) $\hat{\Gamma}$ behavior is supposed to be formed under the influence of the electron-electron correlations. As a result, vertex function $\hat{\Gamma}$ and GF anomalous part are supposed to have the well known d-type behavior [2] in cuprates. $g_{Rnod}$ is the nodal part of a retarded electron Green function, $\alpha_{nod}^2 F$ is the nodal part of the electron-phonon interaction spectral function, $\Gamma N_{0nod}(\xi)$ represents the bare (not renormalized with the EP interaction) variable fraction of the density of states defined by the following expression:

$$\int_{S(\xi)} \Gamma(\boldsymbol{p}_{nod}, \boldsymbol{p}', \omega') \frac{d^2 \boldsymbol{p}'}{v_{\xi \boldsymbol{p}'}} d\xi = \int_{S(\xi)} \Gamma N_{0nod}(\xi, \omega') d\xi$$

at the energy of bare electrons $\xi$ marked from a Fermi level with the electron impulse $\boldsymbol{p}_{nod}$ oriented in the nodal direction. $\Gamma(\boldsymbol{p}, \boldsymbol{p}', \omega')$ is the coulomb vertex being greatly reduced near the optimal doping value $\delta$ due to the forward scattering peak (FSP) in the processes of the correlated electron scattering [11] for the transverse impulse component $q_\perp > \delta \cdot \pi / a$ with $a$ as a translation constant of the cuprates plane. As a result we obtain the separated equations for the nodal and antinodal Green functions in the neighborhood of the optimal doping $\delta \approx 0.16$ [8]. Let us neglect the dependence of $\alpha_\varphi^2 F$ on $\xi, \xi'$ values. We shall replace the mass renormalization factor $Z(\vec{p}', \omega)$ with the quantity $Z_\varphi(\xi, \omega)$ corresponding to the constant energy $\xi$ in the direction defined with the angle $\varphi$. For the $(1,1)$ component of the matrix SE nodal imaginary part $\mathrm{Im}\Sigma_\varphi(\omega) = -\mathrm{Im}Z_\varphi(\omega)\omega + \mathrm{Im}\chi_\varphi(\omega)$ the following expression should be obtained:

$$\mathrm{Im}\Sigma_{nod}(\omega) = -\pi\int_0^{+\infty} dz \alpha_{nod}^2(z)F(z) \times$$
$$\times\{[N_{nod}(\omega-z) + N_{nod}(\omega+z)]n_B(z) + N_{nod}(\omega-z)f(z-\omega) + N_{nod}(\omega+z)f(z+\omega)\}. \quad (1)$$

In (1) $n_B(z)$ is the Bose distribution function, $f(z')$ is the Fermi distribution function, $cth(\omega_{ph}/T) \approx 1$. In (1) the renormalized with the EP interaction nodal part of the density of the electron states $N_{nod}(z')$ is expressed through the "bare" partial nodal density of electron states

$N_{nod}(z') = -\dfrac{1}{\pi}\int_{-\mu}^{\infty} d\xi' \Gamma N_{0nod}(\xi')\,\mathrm{Im}\, g_{Rnod}(\xi',z')$. In the following the subscript "nod" shall be omitted or replaced for brevity with the subscript "n". We shall further approximate the expression for the bare (modified with the doping but not with the EP interaction) non constant electron density of states $\Gamma N_0(\xi)$ with an abrupt drop in the energy $\xi$ as follows: $\Gamma N_0(\xi) = \varsigma \cdot \Gamma N_0$ at

$-W[1+(\varsigma-1)d]/\varsigma \leq \xi \leq -d\cdot W$ and $\Gamma N_0(\xi) = \Gamma N_0$ at $-d\cdot W < \xi \leq W$ (Fig.1) where $\left(1+\dfrac{1}{\varsigma}\right)W$ is the width of the initial bare band, d=0.16 is the optimal level of doping, $\varsigma \approx 1.38$ [12,13]. Such a model of the variable electron density of states leaves the zone half filled in the absence of the doping. Of course, this model density of states acquires physical meaning only at the doping levels of at least 0.1. We will consider for simplicity that the hole or electron type doping shifts the chemical potential $\mu$ from the half filling value in a linear mode as a function of a doping degree $p$. In Fig.1 the overdoped bare model density of states is compared with the calculated [13] density of states. At the optimal level of the hole-type doping an abrupt drop in the electron density of states will be located directly at the Fermi level. As is easily seen from (1), this circumstance will lead to a sharp drop in the electron decay rate $\text{Im}\Sigma_{nod}(\omega)$ at the optimum level of the doping. Such an effect is easily explained with the inability of the electron scattering to the occupied at low temperatures states below the Fermi energy, and the low probability of the electron scattering to the free states, which lie above the Fermi surface, since the density of states above the Fermi surface at the optimal hole doping level is dramatically reduced. Neglecting the dependence of $\text{Re}\Sigma_\varphi(\xi',\omega)$ as well as dependence of $\text{Im}\Sigma_\varphi(\xi',\omega)$, $\text{Re}\varphi(\xi',\omega)$, $\text{Im}\varphi(\xi',\omega)$ values on $\xi'$ and taking into account the standard expression for the retarded GF in the Nambu representation we obtain as a result of a direct integration on $\xi'$ after performance of the analytical continuation $i\omega_p \to \omega + i\delta$ the following equation for the real part of the complex anomalous **nodal** order parameter fraction near $T_c$:

$$|Z(\omega)|\text{Re}\Delta(\omega) = P\int_{-\infty}^{+\infty} dz' \Gamma N_0 \cdot K^{ph}(z',\omega) \frac{1}{z'|Z(z')|} \{\Xi(z')[\text{Re}\Delta(z')\text{Re}Z(z') + \text{Im}\Delta(z')\text{Im}Z(z')] +$$
$$+ 0.5\Lambda(z')[\text{Im}\Delta(z')\text{Re}Z(z') - \text{Re}\Delta(z')\text{Im}Z(z')]\} \qquad (2)$$

For the imaginary part $\text{Im}\Delta(\omega)$ of the nodal order parameter fraction after straightforward integration on $\xi'$ we obtain at $cth(z/2T) \approx 1$

$$\text{Im}\Delta(\omega)|Z(\omega)| = \pi\int_{-\infty}^{+\infty} dz \frac{1}{2}\Gamma N_0 \cdot \alpha^2(|\omega-z|)F(|\omega-z|)sign(\omega-z)\frac{1+th\frac{z}{2T}}{z|Z(z)|}\{\Xi(z)[\text{Re}\Delta(z)\times \qquad (3)$$
$$\times \text{Re}Z(z) + \text{Im}\Delta(z)\text{Im}Z(z)] + 0.5\Lambda(z')[\text{Im}\Delta(z')\text{Re}Z(z') - \text{Re}\Delta(z')\text{Im}Z(z')]\},$$

where $K^{ph}(z',\omega)$ is a standard kernel appearing in EE. Therefore, it is clear that the right hand side of the finite zone width EE should be represented with the linear superposition of the real $\text{Re}\Delta(z)$ and imaginary $\text{Im}\Delta(z)$ parts of the order parameter.

Here $\Delta(\omega) = \varphi(\omega)/|Z(\omega)|$, $|Z(z')| = (\text{Re}^2 Z(z') + \text{Im}^2 Z(z'))^{\frac{1}{2}}$, an integral on $z'$ in (2) is taken in the sense of a principal value, indicated with the symbol P. We shall assume expressions (2, 3) to be averaged over the angle $\varphi$ within the nodal region. The expressions $\Xi(z')$ and $\Lambda(z')$ represent the factors that take into account the finiteness of the electron band width and the variable electron density of states

$$\Xi(z) = \arctan\left[\frac{\xi_H + z'\text{Re}\,Z(z',p)}{-\text{Im}\,\Sigma(-z',p)}\right] + \arctan\left[\frac{\xi_H - z'\text{Re}\,Z(z',p)}{-\text{Im}\,\Sigma(z',p)}\right] + $$
$$+\varsigma\arctan\left[\frac{-\xi_L - z'\text{Re}\,Z(z',p)}{-\text{Im}\,\Sigma(-z',p)}\right] + \varsigma\arctan\left[\frac{-\xi_L + z'\text{Re}\,Z(z',p)}{-\text{Im}\,\Sigma(z',p)}\right] > 0 \quad (4)$$

where $\xi_H = W - \delta\mu + \text{Re}\,\chi(z')$, $\xi_L = -W/\varsigma - \delta\mu + \text{Re}\,\chi(z')$. The imaginary part of the SE $-\text{Im}\,\Sigma(\pm z')$ in (2), (3) never becomes zero and remains finite even on the Fermi surface. The EP interaction constant is evaluated [9] as $\lambda \sim 1.21$ as the slope of $\text{Re}\,\Sigma(\omega) \sim -\lambda\omega$ at low frequency, which is found to be in consent with the results presented in the reviews [4,5].

## 2 High $T_c$ value in cuprates as a consequence of the electron density of states behavior

The calculations for the $\text{Re}\,Z(\omega,p,T)$, $\text{Im}\,Z(\omega,p,T)$, $\text{Re}\,\chi(\omega,p,T)$ and $\text{Im}\,\chi(\omega,p,T)$ values contained in (2) - (3) have been implemented within the formalism developed in [9] using a zone width value W=5 [12,13] in the dimensionless units expressed in the characteristic phonon energy $\omega_D$. The finite zone factor $\Xi(z)$ behavior which takes into account the electron-hole nonequivalence is shown in Fig.2. The logarithmic factor $\Lambda(z)$ has significantly less influence on $T_c$ than the finite zone width factor $\Xi(z)$ (4) because of its smallness at low frequency (Fig.3.), therefore, for brevity, we shall not give its explicit form. From Figs.2, 3 it becomes clear that the negative contribution to the $\text{Re}\Delta(\omega<\omega_D)$ and, consequently to the $T_c$, is greatly depressed, and at $\omega>W$ is completely "cut off". As is evident from the behavior of the used electron density of states, the effect of the high-frequency suppression is significantly weaker as expressed in the case of the electron type doping, which leads to the low $T_c$ in n-type doped cuprates. The behavior of both factors (Figs.2, 3.) also shows that at the electron-type doping the high-frequency contribution to the kernel of the EE, which reduces $T_c$, is much higher than at the p-type doping. The sharp weakening of the electron decay rate $\text{Im}\Sigma_{nod}(\omega)$ at the optimal hole doping, as described above, leads to the fact that the EP interaction potential at the optimal p-type doping at frequency not exceeding the maximum phonon frequency $\omega_D$ is weakened very slightly, as the electron damping $\text{Im}\Sigma_{nod}(\omega)$ appears in the denominator of the arctangent in (4), determining the high frequency suppression effect in the EE. This fact, together with the finite width of the electron band is a second critical factor influencing the emergence of the high $T_c$ values in cuprates. At high carrier concentration the electron-electron interaction begins to dominate in the formation of the electron inverse lifetime $\text{Im}\Sigma_{nod}(\omega)$. The iterative solution of the set of equations (2), (3) is performed for clarity with the use of Einstein-type spectral function of the EP interaction $\alpha^2F(z)=\lambda\omega_0\delta(z-\omega_0)/2$ which "models" the experimentally observed Eliashberg function $\alpha^2F(z)$ [8] for Bi2212 with the average phonon frequency $\omega_0 \approx 60$ meV. In $T_c$ calculations in the cuprates, we proceed from the experimental fact that $T_c$ is determined only by the nodal GF [14,15]. The order parameter real part $\text{Re}\Delta_{nod}(\omega=0.3,T)$ on T behavior in the case of the $\delta=0.16$ hole-type optimal doping with the resulting value $T_c \sim 0.28\omega_0$ is presented on Fig.4. Both $\text{Re}\Delta_{nod}(\omega,T)$ and $\text{Im}\Delta_{nod}(\omega,T)$ vanish at $T_c$ for all frequencies at once, so the frequency $\omega = 0.3$ is chosen at random. The terms in equations (2, 3) for the order parameter proportional to the inverse electron lifetime $\operatorname{Im} Z(z')$ with the further increase in the doping degree lead to a decrease of the $T_c$ value. At the larger electron concentration the electron-electron interactions should lead to the much larger value of $\operatorname{Im} Z_{el-el}(z')$ and therefore to the complete suppression of the high-$T_c$ superconductivity effect. A similar calculation for the electron-type doping $\delta = 0.16$ results with $T_c = 0.14\omega_0$ value.

### 3. Conclusions

1. It is found that the high frequency suppression in the newly derived Eliashberg equations for the finite zone width EP system together with the abrupt fall in the electron density of states above the Fermi surface at the optimal dagree of the hole-type doping are the crucial factors for the high temperature superconductivity (HTS) phenomenon. When substituting $\omega_0$ values into the $T_c$ formula for the $\omega_0 = 40\ meV$ we get a bit inflated $T_c \sim 123K$ value compared with the experimental value $T_c = 95K$ in Bi2212, which is explained by the neglect in our calculations of the Coulomb pseudopotential and by the $T_c$ overestimation in the Einstein model.

2. The $T_c$ value in the electron-doped cuprates seems to be quite ordinary and should be calculated in a standard manner [1-6].

3. The $\operatorname{Re}\Delta(\omega)$ increase at $0 < \omega < \omega_D$ due to a sharp weakening of the negative $\operatorname{Re}\Delta(\omega > \omega_D)$ contribution is a main factor in the $T_c$ increase in the cuprates. This property of the EE had not been discovered in the formalism of the previous works [1-6].

4. In the existing work [10] devoted to the analysis of the finite electron zone width EE the density of the electron states in contrast to our work was assumed to be constant.

5. The accounting for the electron pseudopotential leads to only a slight decrease in the calculated $T_c$ value. All obtained results are also valid for the standard EP system not manifesting d-properties.

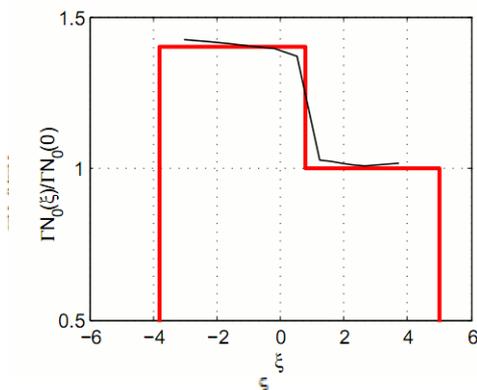

FIG.1. Relative bare model electron density of states in hole-type overdoped p=0.24 Bi2212 as a function of the electron energy ξ, represented in dimensionless $\omega_D$ units (thick line), calculated [13] electron density of states (thin line)

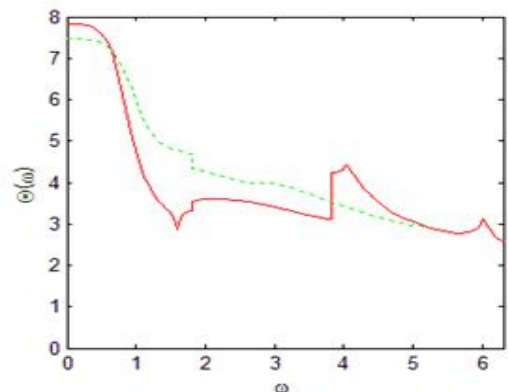

Fig.2 The dimensionless finite zone width factor Θ(ω) for the optimal p=0.16 hole doping (solid line) and for the same degree of the electron doping (dotted line) for Bi2212 at T=0.25

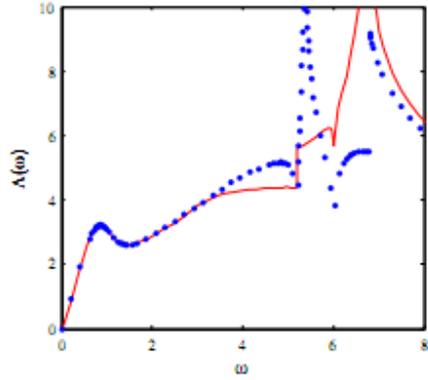

Fig.3. Dimensionless logarithmic factor $\Lambda(\omega)$ for optimal p=0.16 hole doping (solid line) and for the same degree of the electron doping (dotted line) for 2212 at T=0.15

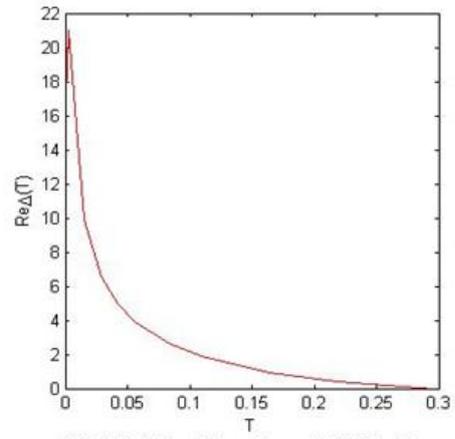

FIG.4. $Re\Delta(T)$ on T dependence in Bi2212 at the hole type p=0.16 doping. T,$\Delta$ in averaged phonon frequency $\omega_D$ units